# Low cost electrical probe station using etched Tungsten nanoprobes: Role of cathode geometry

Rakesh K. Prasad[a] and Dilip K. Singh[a*]

[a]Department of Physics, Birla Institute of Technology Mesra, Ranchi-835215.
*Email: dilipsinghnano1@gmail.com

## Abstract

Electrical measurement of nano-scale devices and structures requires skills and hardware to make nano-contacts. Such measurements have been difficult for number of laboratories due to cost of probe station and nano-probes. In the present work, we have demonstrated possibility of assembling low cost probe station using USB microscope (US $ 30) coupled with in-house developed probe station. We have explored the effect of shape of etching electrodes on the geometry of the microprobes developed. The variation in the geometry of copper wire electrode is observed to affect the probe length (0.58 mm to 2.15 mm) and its half cone angle (1.4° to 8.8°). These developed probes were used to make contact on micro patterned metal films and was used for electrical measurement along with semiconductor parameter analyzer. These probes show low contact resistance (~ 4  ) and follows Ohmic behavior. Such probes can be used for laboratories involved in teaching and multidisciplinary research activities and Atomic Force Microscopy.

*Keywords:  electrochemical etching, tungsten tip, DC voltage, Low cost probe station.*

## I. INTRODUCTION

Advancement in the field of nanofabrication has led to miniaturization of devices to nanometers. Research labs and teaching efforts in the field of electronics and opto-electronic devices to such small dimensions, require probes for micron or smaller size. Additionally, these factors have limited the access of experts from various domains of science and engineering to explore nanoscale structures for multi-disciplinary applications. Various research groups have attempted to devise methods of fabricating metallic nano-probes using cost effective techniques to achieve reproducible tip geometry. There are various methods for the formation of tungsten tip like cutting **(1, 2)**, mechanical pulling **(3-11)**, grinding **(12, 13)**, ion milling **(14-18)**, ion beam–induced deposition **(19)**, electrochemical etching **(20-35)** and electrochemical machining **(36)**. Recently in 2019 Yamaguchi et al. introduced a new method called flame etching to fabricate tungsten tip. **(37)** In 1951, Miller et al. reported about possibility of fabrication of sharp metal tips by electrochemical process **(38).** With time there have been refinements in the methods to get sharp, smooth and long taper tip with perfectly conical geometry.

The driving force for the research in this direction has been the concern about reproducibility of probe geometry and their immense application in nano characterization tools for topography, electrical and optical measurements **(34, 39, 40).** Few notable improved techniques for electrochemical etching are drop-off methods with direct current (DC) voltage **(20, 41)** dynamic etching technique **(39),** reverse chemical etching **(24, 42, 43)**. Chemical etching is one of the most effective methods for fabricating various types of nano-probes with different geometry. For the purpose of chemical etching, Sodium hydroxide (NaOH) or Potassium hydroxide (KOH) as electrolytes are used with varying molar concentration in the range 0.1 M -10 M **(26, 27, 29, 30, 44).** Tungsten wire was used as an anode during etching while a variety of materials like stainless steel **(25),** chromium- nickel stainless steel **(44),** iridium **(26)**, platinum **(28)** and tungsten wire **(29)** have been used as cathode**.** Although different researchers have used cathode of varying geometry (wire or rod, circular loop, ring and L-shaped etc) **(26, 28, 29)** there has not been any study on the effect of cathode geometry on the resultant nano-probes formed. There are various parameter which affect the rate of electrochemical etching, formation of probe and their geometry like electrolyte concentration, immersion depth of wire, size and position of cathode, applying DC or Alternating current (AC) varying voltages and diameter of tungsten wire **(45)**. Applying AC voltage is usually more difficult to control and it affects the smooth formation of probes, in comparison to DC voltages **(46)**. The meniscus formed between the wire-electrolyte interface due to capillary forces directly influences the probe-tip shape **(44)**. Low DC voltage during etching was suggested by Ibe et al. to give less oxide film formation as in comparison with the use of high voltage **(47)**. Till now, the initial tungsten wire diameter for nano-probe formation has been investigated to vary from 50 µm to 250 µm **(28-30, 34, 37, 39)**, however a very few have used higher diameters such as 1000 µm. **(35, 44)** Most groups have used drop-off method with some modification in the circuit for automatic cut-off in the power supply after fall of the tungsten wire into the solution, **(25, 35, 44)**, even though it is more complicated and expensive.

Nowadays for fine fabrication of nano-probes the complete electrochemical etching is sub-divided into a few step processes involving certain parameter changes. Two step etching process involving coarse and fine etching was demonstrated by Olivier L. Guise et,al in 2002, where the DC voltage was changed at each step **(26)**. In 2012 Yasser khan et al. introduced a two dynamic electrochemical etching in which simple etching of the probe was followed by a secondary step of cleaning, drying and re-installing the probe but in a smaller immersion depth than in the first etching **(30)**. Recently in 2018 Danish Hussain et al. used a trapezoidal potential for micro needle formation followed by a DC potential for further probe tip formation **(34)**. Shanli Qin in 2019 also followed a two-step technique and fabricated Nano-probes tips **(35)** .

Cleaning is essential after the fabrication mechanism of nano-probes to avoid the development of a thin oxide layer of $H_2O$, $H_2$, $O_2$, $CO$, and other hydrocarbons **(48, 49)** due to dissolution of tungstate anions ($WO_4^{2-}$) in water during the growth **(50)**. These oxide films increase the thickness of the probes and reduce their performance as a good electrical contact for electrical measurements **(25)**. Different techniques are used to reduce or remove the oxide layer including annealing inside ultra-high vacuum scanning tunneling microscopy chamber **(25)**, chemical cleaning by Hydrofluoric acid **(51)**, immersing in deionized water, acetone, ethanol or methanol **(26).** Sometimes post-cleaning, the tip is dried gently either by clean nitrogen gas or by sputtering **(25)** .

In our study, we have explained the role of electrodes with different shapes and reproducibility of sharp, smooth and long tapered tip by using different geometrical pattern of copper wire that is used as another electrode. Using a two-step method, we etched the tungsten tip by NaOH solution for one hour, followed by further etching in KOH solution till the etching process is completed. We have also explored the effect of cathode geometry on the formation of sharp and smooth probes. Formation of tungsten tip is then analyzed by the optical microscope and field emission scanning electron microscope (FESEM). These nano-probes were used in developing a low-cost, effective probe-station that can be utilized for electrical measurements in micro-electronics.

## II. EXPERIMENTAL DETAILS

For the electrochemical etching process to fabricate metallic nano-probes, we used 2M NaOH and 2M KOH solution sequentially during two step etching process. During etching copper wire of diameter 0.3 mm was used as cathode while tungsten wire of diameter 1.0 mm (purity 99.9 %, Sigma Aldrich) was used as anode. DC power supply of 5V (Scientech, Model 4073) was used as driving source for etching process. Etching was performed using cathode of varying shape: straight, circular, triangular, square and pentagon folded wire. Two-step process with different etchant solution was taken due to difference in their etching rates. Etching with NaOH was first performed (referred as $1^{st}$ step) and followed by etching with KOH ($2^{nd}$ step). After etching with NaOH, the tungsten wire was cleaned by ethanol and de-ionized water. During first step, etching with 5 volts DC offers optimum etching rate resulting into smooth outer surface of the tips, while etching at higher voltages like 8 volts results into relatively much faster etching rate with irregular rough surface. NaOH leads to faster etching rate as compared to KOH, but results into rough outer surface. Initially etching with NaOH was performed to reduce the wire diameter to few mm (step 1) and subsequently the wire was etched with KOH solution with 3 volts DC power to smoothen the outer surface as the outer diameter reduces to ~ 50 nanometers (step 2). Further we explored the effect of copper cathode geometry on the etching rate and roughness of outer surface of formed nano probes. For comparison of resultant geometries during the etching process other parameter like separation between two electrodes (2 cms) and portion of the tungsten rod dip inside the solution (3 cms) along with DC voltage was kept fixed. For anode with varying geometrical shape the sides of each shape was kept constant 3 cms (triangular shape, rectangular shape and pentagon), while for circular shape the diameter was kept at 3 cms. Visibly the tungsten anode in the form of straight wire was located at the centre of the copper cathode of various geometries. The surface morphology and the diameter and length of probes were analysed using a high–resolution ZEISS sigma 300 field emission scanning microscope (FESEM). During the process of electrochemical etching, following reaction takes place: **(47)**

$$6H_2O + 6e^- \rightarrow 3 H_2 (g) + 6 OH^- \quad \text{On cathode}$$

$$\underline{W(s) + 8 OH^- \rightarrow WO_4^{2-} + 4 H_2O + 6e^- \quad \text{On Anode}}$$

$$W(s) + 2 OH^- + 2 H_2O \rightarrow WO_4^{2-} + 3 H_2 (g) \text{ (over all reaction)}$$

Etching with NaOH solution was performed for 1 hour and after that etching was done using KOH solution till the etched portion of the wire drop-off known as drop off method. The entire process was monitored by digital microscope to elucidate the effect of minuses formed on the neck formation on the wire's outer surface leading to formation of tip.

**Result and discussion**

Figure 1(a) shows the schematic of the experimental setup used for fabrication of metallic nano-probes. Figure 1(b) shows the photograph of the etched anode with NaOH for 01 hour. Etching leads to formation of neck at the air-solvent interface, schematically as shown in figure 1(c). Figure 1(d) shows the photograph of tip formed before drop-off. During etching a meniscus is formed around the anode after application of DC voltage, clearly visible with naked eye. Figure 1(d) and 1(e) shows the photographs of process of electrochemical etching using anodes of circular and square geometry respectively, progress of etching process leads to formation of bubble at the cathode surface as seen in figure 1(d) and 1(e). Etching with anode of different geometry results into difference in shape of etched neck of varying sizes as estimated using Vernier caliper (table-1).

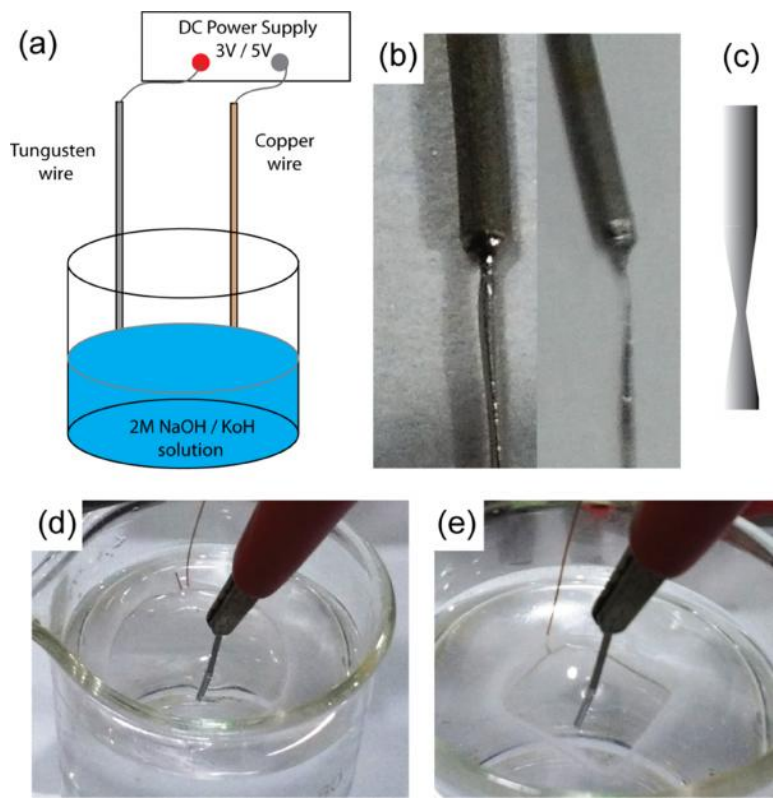

Figure.1 (a) shows schematic of the experimental setup used for electrochemical etching of Tungsten wire to fabricate nanoprobes. (b) Shows photographs of formation of neck during etching at air-solvent interface. (c) Shows schematically diagram of formation of neck during etching. (d) and (e) shows photograph of etching using circular and square cathodes. Formation of meniscus is clearly visible around the anode resulting into neck creation on the tungsten wire surface.

Table-1: Effect of anode geometry on the neck formation during electrochemical etching.

| Shape of electrode | Area (cm$^2$) | Top (mm) | Neck width (mm) | Bottom (mm) |
| --- | --- | --- | --- | --- |
| *Straight* | 3.00 | 0.42 | 0.30 | 0.35 |
| *Circular* | 7.07 | 0.62 | 0.59 | 0.50 |
| *Triangular* | 3.90 | 0.36 | 0.40 | 0.47 |
| *Square* | 9.00 | 0.40 | 0.21 | 0.37 |
| *Pentagon* | 15.48 | 0.10 | 0.20 | 0.40 |

Further continued etching using NaOH results into drop-off of etched portion after 90 minutes. A closer look using FESEM indicates rough surfaces at the probe tip, making them unsuitable to be used as electrical contact probe and Atomic force microscopy probes. Prolonged etching using NaOH alone leads to drop-off for all the cathode geometry and results into rough outer surface of the tip edge (see in table-S1, supplementary information). Figure 2 shows the relative effect of etching with NaOH and KOH separately using triangular geometry of cathode. Figure 2(a) and (b) shows etched probe using NaOH at two different magnifications, while figure. 2(c) and (d) shows etching with KOH alone. Etching with KOH results into smooth outer surface with much smaller probe length.

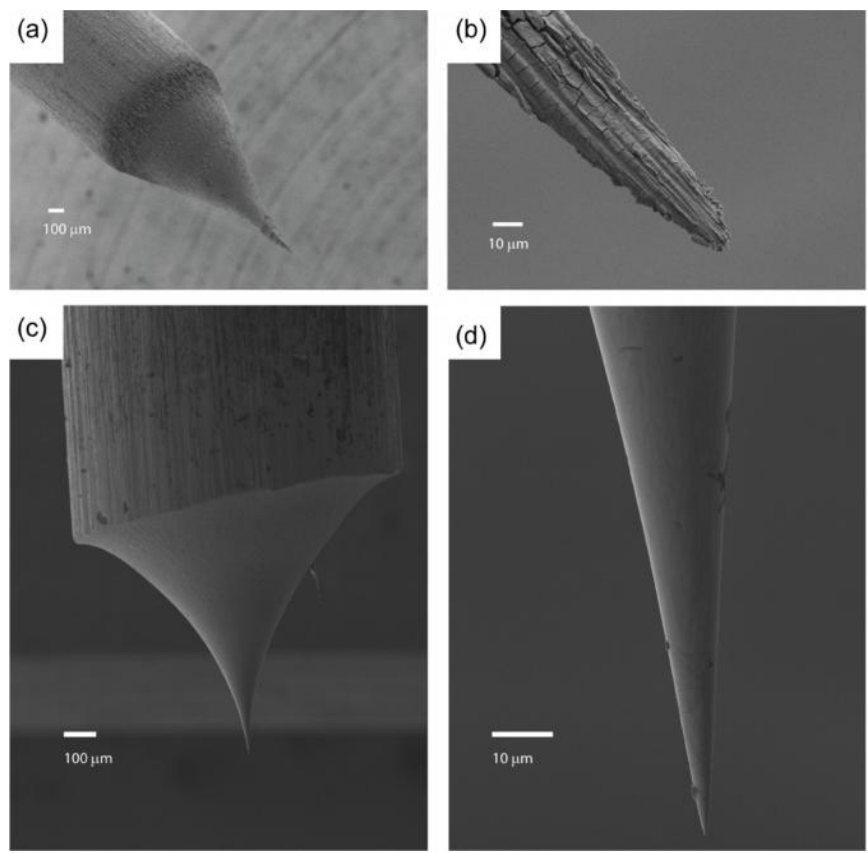

Figure.2 shows relative effect of etching with NaOH and KOH. (a) & (b) shows etching of tungsten wire using NaOH only for about 90 minutes at two different magnifications. (c) & (d) shows etching using KOH only for 100 minutes. KOH etching results into smooth surface.

Since NaOH based etching occurs at much faster rate but results into undesirable outer surface of the probes, while etching using KOH gives rise to smooth surface with much slower rate. A combined process of etching first with NaOH for 01 hour followed by etching with KOH till drop-off results into probes with tip radius in nanometer range and smooth outer surface.

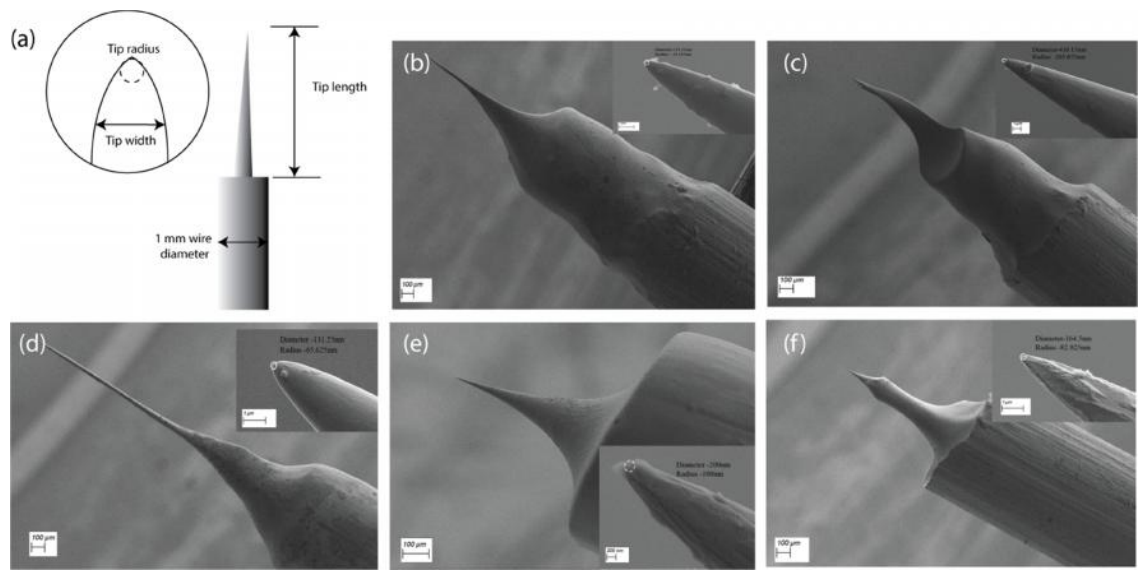

Figure 3 shows the effect of cathode geometry on the shape and tip radius of fabricated nano-probes. The tip length and tip radius was estimated schematically as shown in the figure (a). Probes fabricated under similar conditions with two step process and cathodes of different geometry are shown in figure (b) straight electrode, (c) circular electrode, (d) triangular electrode, (e) square electrode and (f) pentagon electrode. Inset of the fig. (b)-(f) shows the apex of the fabricated respective probes.

Above figure shows the FESEM images of probes etched with two step process and using cathodes of different geometry. Figure 3(a) shows the schematic of estimation of tip radius and probe length.

The FESEM images of probes fabricated using two-step process under similar conditions using cathodes of varying geometry using straight, circular, triangular, square and pentagon electrode are shown figure 3 (b)- (f) respectively. The estimate values of tip radius and probe length are summarized in table-II.

Table-II The geometrical parameters of probes obtained with electrodes with varying geometry.

| Shape Of cathode | Length (mm) | Half angle (degrees) | Radius of tip (nm) |
|---|---|---|---|
| Straight | 2.05 | 3.3 | 55 |
| Triangular | 0.58 | 5.0 | 100 |
| Circular | 1.80 | 4.8 | 205 |
| Square | 2.15 | 1.4 | 66 |
| Pentagon | 0.96 | 8.3 | 82 |

Interestingly etching with straight and square electrode gives rise to smallest tip radius of ~ 60 nm followed by triangular and circular cathodes resulting into tip radius of 100 nm and 205 nm respectively. Triagular electrode results into smallest probe length (0.58 μm) while, square electrode results into probes with length 2.15 mm. A closer look into the half angles of the probes formed shows etching with square electrode results into sharpest probe, while etching with triangular electrode results into probe with half angle of 5°. Among all the cathode shapes, the best probe tip shape results from triangular electrode. Triangular electrode results into probes with symmetric shape with single mode, while other shapes of electrodes into multimode structure, which may give rise to multiple resonance frequencies for their end use as scanning probe microscopy. The eteching with varius cathode geometry were repeated before summarizing the shape of resultant probes.

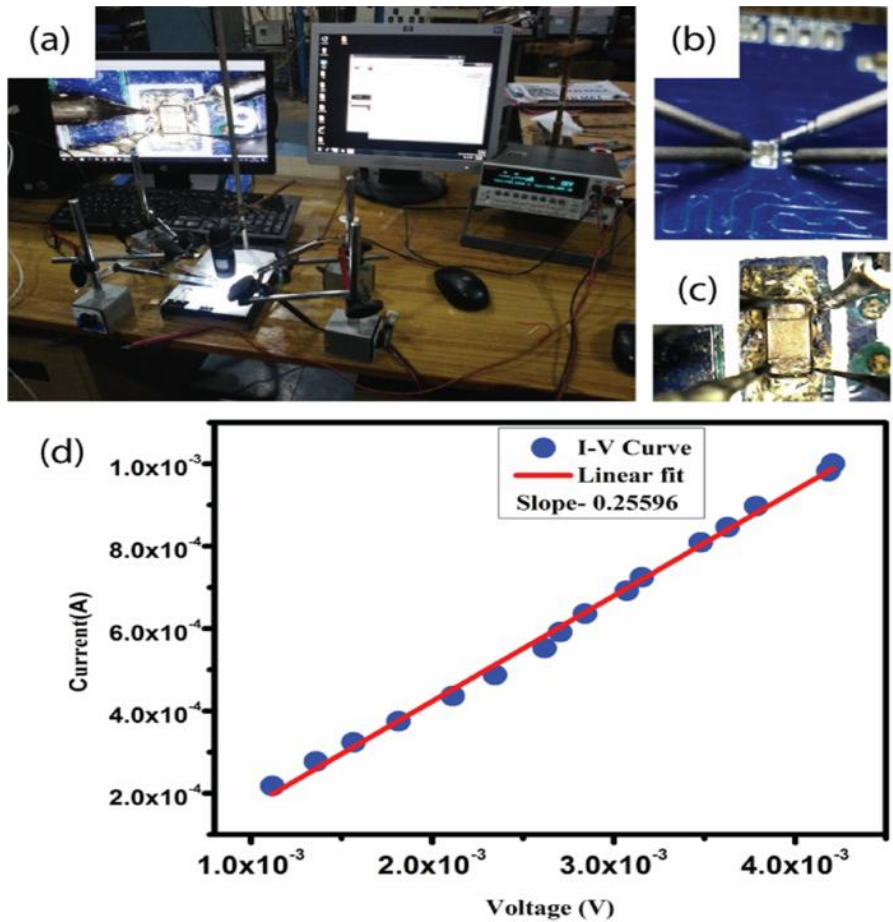

Figure 4: (a) Shows photograph of developed low cost probe station. (b) & (c) shows contact made over one of the connecting pads of a portion of RAM using devloped probes at lower and higher resolutions. (d) I-V measurement for a metallic strip of micrometer size by using devloped probe- station.

A nano-probe station with the benefit of minimalistic contact area for micro and nano electrical characterisation was established using the developed nano-probes and a low cost USB based microscope with a magnification upto 1000×. The nano-probe station (as shown in figure 4(a)) was assembled with the help of four screw gauges, each mounted on a magnetic stand, that aids in the fine mechanical movement of each nano-probe. The nano-probes that are linked with the screw gauge, have an intermediatery insulating pipe to avoid any type of metallic contact. Then, each nano-probe end is coupled to a single wire connector that is utilised in giving an electrical response that is determined using Keitheley 2410 SourceMeter. An approximate visualisation of the contact between metallic sample strip with the nano-probe is determined by using a digital microscope. This nano-probe station is employed to determine the I-V characterisitics of a micrometer size metallic strip as shown in Figure 4a. The linear fit of the electrical response gave us the resistance of the metallic strip to be 3.8 ohms.

**Conclusion**

Etching with NaOH results in rough probe surface and results in probe apex size of ~ 2 μm range, whereas etching with KOH results in shorter probe length with smoother surface and probe apex size of ~ 500 nm range. There has been no report about the formation of nanoprobes (probes with tip apex size of nm) using direct etching with DC sources, without any additional circuit breaker or modification in circuit. As such, the combined process of sequential etching using NaOH followed by KOH results in the formation of nanoprobes using DC source alone making it inexpensive. For the first time, geometry of the cathode used for etching is found to affect the probe apex angle and probe length. Cathode of triangular geometry is found to form sharp probe- tip with single mode. Multiple groove on the tip edge results into multimode giving rise to multiple resonance frequencies: which is not desirable for probe spectroscopy and microscopy measurements. Using the developed nanoprobes the possibility to use them as low cost probe station for nanoelectronics has been demonstrated.

**Acknowledgments**

One of the authors Rakesh K. Prasad thanks TEQIP-III for fellowship. Dilip K. Singh thanks DST, Government of India (Fellowship IFA13-PH65) and Seed Money Scheme, Birla Institute of Technology, Mesra for funding.

**Data Availability**

The data that support the findings of this study are available from the corresponding author upon reasonable request.

**Reference**

1. Garnaes J, Kragh F, Mo/rch KA, Thölén AR. 1990 Transmission electron microscopy of scanning tunneling tips *Journal of Vacuum Science & Technology A: Vacuum, Surfaces, and Films* **8(1)** 441-4

2. Gorbunov AA, Wolf B, Edelmann J. 1993 The Use of Silver Tips in Scanning-Tunneling-Microscopy *Review of Scientific Instruments* **64(8)** 2393-4

3. Lazarev A, Fang N, Luo Q, Zhang X. 2003 Formation of fine near-field scanning optical microscopy tips. Part II. By laser-heated pulling and bending *Review of scientific instruments* **74(8)** 3684-8

4. Betzig E, Trautman J, Harris T, Weiner J, Kostelak R. 1991 Breaking the diffraction barrier: optical microscopy on a nanometric scale *Science* **251(5000)** 1468-70

5. Yakobson B, Moyer P, Paesler M. 1993 Kinetic limits for sensing tip morphology in near-field scanning optical microscopes *Journal of applied physics* **73(11)** 7984-6

6. Valaskovic G, Holton M, Morrison G. 1995 Parameter control, characterization, and optimization in the fabrication of optical fiber near-field probes Applied optics **34(7)** 1215-28

7. Xiao M, Nieto J, Machorro R, Siqueiros J, Escamilla H. 1997 Fabrication of probe tips for reflection scanning near-field optical microscopes: Chemical etching and heating-pulling methods *Journal of Vacuum Science & Technology B: Microelectronics and Nanometer Structures Processing, Measurement, and Phenomena* 15(4) 1516-20

8. Essaidi N, Chen Y, Kottler V, Cambril E, Mayeux C, Ronarch N, et al. 1998 Fabrication and characterization of optical-fiber nanoprobes for scanning near-field optical microscopy *Applied optics* **37(4)** 609-15

9. Hibi T 1956 Pointed filament (I) its production and its applications *Microscopy* **4(1)**10-5

10. HIBI T, ISHIKAWA K 1960 Study of Point Cathode by Using Müller's Type Microscope *Microscopy* 9(2):13-5

11. Binnig G, Rohrer H, Gerber C, Weibel E 1982 Surface studies by scanning tunneling microscopy. Physical review letters **49(1)**57

12. Nonnenmacher M, o'Boyle M, Wickramasinghe HK 1991 Kelvin probe force microscopy *Applied physics letters* **58(25)** 2921-3

13. Held T, Emonin S, Marti O, Hollricher O 2000 Method to produce high-resolution scanning near-field optical microscope probes by beveling optical fibers *Review of Scientific Instruments* **71(8)** 3118-22

14. Biegelsen D, Ponce F, Tramontana J, Koch S 1987 Ion milled tips for scanning tunneling microscopy *Applied physics letters* **50(11)** 696-8

15. Biegelsen D, Ponce F, Tramontana J 1989 Simple ion milling preparation of <111> tungsten tips *Applied physics letters* **54(13)** 1223-5

16. Hoffrogge P, Kopf H, Reichelt R 2001 Nanostructuring of tips for scanning probe microscopy by ion sputtering: Control of the apex ratio and the tip radius *Journal of Applied Physics* **90(10)** 5322-7

17. Akiyama K, Eguchi T, An T, Fujikawa Y, Yamada-Takamura Y, Sakurai T, et al. 2005 Development of a metal–tip cantilever for noncontact atomic force microscopy *Review of scientific instruments* **76(3)** 033705

18. Burke M, Sieloff D, Brenner S 1986 A Combined TEM/FIM Examination of Field Emission as a FIM Specimen Preparation Technique *Le Journal de Physique Colloques.* **47(C7)** :C7-459-C7-62.

19. Onishi K, Guo H, Nagano S, Fujita D 2014 High aspect ratio AFM Probe processing by helium-ion-beam induced deposition *Microscopy* **63**(suppl_1):i30-i

20. Klein M, Schwitzgebel G 1997 An improved lamellae drop-off technique for sharp tip preparation in scanning tunneling microscopy *Review of scientific instruments* **68(8)** 3099-103

21. Lindahl J 1998 Easy and reproducible method for making sharp tips of Pt/Ir *Journal of Vacuum Science & Technology B: Microelectronics and Nanometer Structures* **16(6)** 3077

22. Song JP, Pryds NH, Glejbo/l K, Mo/rch KA, Thölén AR, Christensen LN 1993 A development in the preparation of sharp scanning tunneling microscopy tips *Review of Scientific Instruments* **64(4)** 900-3

23. Muramatsu H, Homma K, Chiba N, Yamamoto N, Egawa A 1999 Dynamic etching method for fabricating a variety of tip shapes in the optical fibre probe of a scanning near-field optical microscope *Journal of microscopy* **194**(Pt 2-3) 383-7

24. Fotino M 1993 Tip sharpening by normal and reverse electrochemical etching *Review of Scientific Instruments* **64(1)**159-67

25. Ekvall I, Wahlstrom E, Claesson D, Olin H, Olsson E 1999 Preparation and characterization of electrochemically etched W tips for STM *Meas Sci Technol.* **10(1)**11-8

26. Guise OL, Ahner JW, Jung M-C, Goughnour PC, Yates JT 2002 Reproducible Electrochemical Etching of Tungsten Probe Tips *Nano Letters* **2(3)**191-3

27. Kulawik M, Nowicki M, Thielsch G, Cramer L, Rust HP, Freund HJ, et al. 2003 A double lamellae dropoff etching procedure for tungsten tips attached to tuning fork atomic force microscopy/scanning tunneling microscopy sensors *Review of Scientific Instruments* **74(2)**1027-30

28. Ju BF, Chen YL, Ge Y 2011 The art of electrochemical etching for preparing tungsten probes with controllable tip profile and characteristic parameters *The Review of scientific instruments* **82(1)** 013707

29. Chang WT, Hwang IS, Chang MT, Lin CY, Hsu WH, Hou JL 2012 Method of electrochemical etching of tungsten tips with controllable profiles *The Review of scientific instruments* **83(8)** 083704

30. Khan Y, Al-Falih H, Zhang Y, Ng TK, Ooi BS 2012 Two-step controllable electrochemical etching of tungsten scanning probe microscopy tips *The Review of scientific instruments* **83(6)** 063708

31. Zhang M, Lian X 2016 Rapid Fabrication of High-Aspect-Ratio Platinum Microprobes by Electrochemical Discharge Etching *Materials* **9**(4)

32. Yang B, Kazuma E, Yokota Y, Kim Y 2018 Fabrication of Sharp Gold Tips by Three-Electrode Electrochemical Etching with High Controllability and Reproducibility *The Journal of Physical Chemistry C* **122**(29) 16950-5

33. Ko J, Jarzembski A, Park K, Lee J 2018 Hydrogel tip attached quartz tuning fork using elastomeric tip mold replicated from electrochemically etched tungsten wire *IEEE Micro Electro Mechanical Systems (MEMS), Belfast* 858-861

34. Hussain D, Song J, Zhang H, Meng X, Xie H 2018 Electrochemical etching of lightweight nanotips for high quality-factor quartz tuning fork force sensor: atomic force microscopy applications *Micro & Nano Letters* **13(8)**1136-40

35. Qin S, Deng H 2019 Electrochemical Etching of Tungsten for Fabrication of Sub-10-nm Tips with a Long Taper and a Large Shank *Nanomanufacturing and Metrology* **2(4)** 235-40

36. Kamaraj AB, Sundaram MM, Mathew R 2012 Ultra high aspect ratio penetrating metal microelectrodes for biomedical applications *Microsystem Technologies* **19(2)**179-86

37. Yamaguchi T, Inami E, Goto Y, Sakai Y, Sasaki S, Ohno T, et al. 2019 Fabrication of tungsten tip probes within 3 s by using flame etching *The Review of scientific instruments* **90(6)** 063701

38. E. W. Miiller 1951 Das Feld ionen mikroskop *Z Physik* 131-136.

39. Hobara R, Yoshimoto S, Hasegawa S, Sakamoto K 2007 Dynamic electrochemical-etching technique for tungsten tips suitable for multi-tip scanning tunneling microscopes *e-Journal of Surface Science and Nanotechnology* **5** 94-8

40. Valencia VA, Thaker AA, Derouin J, Valencia DN, Farber RG, Gebel DA, et al. 2015 Preparation of scanning tunneling microscopy tips using pulsed alternating current etching *Journal of Vacuum Science & Technology A: Vacuum, Surfaces, and Films* **33(2)** 023001

41. Oliva A, Romero G A, Pena J, Anguiano E, Aguilar M 1996 Electrochemical preparation of tungsten tips for a scanning tunneling microscope *Review of scientific instruments* **67(5)** 1917-21

42. Kar A, Gangopadhyay S, Mathur B 2000 A reverse electrochemical floating-layer technique of SPM tip preparation *Meas Sci Technol.* **11(10)** 1426

43. Meza JM, Polesel-Maris J, Lubin C, Thoyer F, Makky A, Ouerghi A, et al 2015 Reverse electrochemical etching method for fabricating ultra-sharp platinum/iridium tips for combined scanning tunneling microscope/atomic force microscope based on a quartz tuning fork *Current Applied Physics* **15(9)** 1015-21

44. Kulakov M, Luzinov I, Kornev KG 2009 Capillary and surface effects in the formation of nanosharp tungsten tips by electropolishing *Langmuir* **25(8)** 4462-8

45. Ju B-F, Chen Y-L, Fu M, Chen Y, Yang Y 2009 Systematic study of electropolishing technique for improving the quality and production reproducibility of tungsten STM probe *Sensors and Actuators A: Physical.* **155(1)**136-44

46. Toh S, Tan H, Lam J, Hsia L, Mai Z 2010 Optimization of AC electrochemical etching for fabricating tungsten nanotips with controlled tip profile *Journal of The Electrochemical Society* **157(1)** E6-E11

47. Ibe J, Bey Jr P, Brandow S, Brizzolara R, Burnham N, DiLella D, et al. 1990 On the electrochemical etching of tips for scanning tunneling microscopy *Journal of Vacuum Science & Technology A: Vacuum, Surfaces, and Films* **8(4)** 3570-5

48. Akama Y, Nishimura E, Sakai As, Murakami H 1990 New scanning tunneling microscopy tip for measuring surface topography *Journal of Vacuum Science & Technology A: Vacuum, Surfaces, and Films* **8(1)** 429-33

49. Lisowski W, Van den Berg A, Kip GA, Hanekamp L 1991 Characterization of tungsten tips for STM by SEM/AES/XPS *Fresenius' journal of analytical chemistry* 341(3-4):196-9.

50. Kelsey GS 1977 The anodic oxidation of tungsten in aqueous base *Journal of the Electrochemical Society* **124(6)** 814-9.

51. Hockett LA, Creager SE 1993 A convenient method for removing surface oxides from tungsten STM tips *Review of scientific instruments* **64(1)** 263-4

# Low cost electrical probe station using etched Tungsten nanoprobes: Role of cathode geometry

Rakesh K. Prasad[a] and Dilip K. Singh[a*]

[a]*Department of Physics, Birla Institute of Technology Mesra, Ranchi-835215.*

*Email: dilipsinghnano1@gmail.com*

When we etched 1.5 hour for $1^{st}$ step by 2M NaOH solution and $2^{nd}$ steps by using 2M KoH solution by further etching we obtain irrugular surface with tip radius in 2.07 µm having length 1.78mm

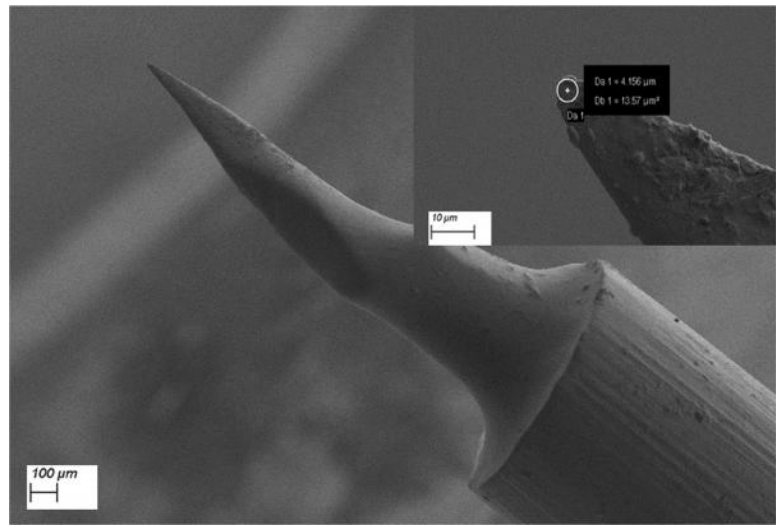

Fig. S1 FESEM image of probe develped by 1step etching for 1.5 hours and further etching with KoH Solution .

Table S1: Result of the etching with different shape of electrode only by one step process but we are able to form micro/submicro tip radius with irrugular surface.

| Shape of electrode | Length (mm) | Half angle (in degree) | Radius of Tip (~m) |
|---|---|---|---|
| *Straight* | 3.04 | 9.4 | 5.12 |
| *Circular* | 1.60 | 8.65 | 2.57 |
| *Triangular* | 0.98 | 45 | 2.85 |
| *Square* | 1.22 | 8.45 | 1.45 |
| *Pentagon* | 0.81 | 21.1 | 1.17 |